\documentclass[11pt]{article}
\usepackage{amssymb}
\begin{document}

\title{Generalized Tu Formula and Hamilton Structures of Fractional Soliton Equation Hierarchy}
\author{Guo-cheng Wu$^{1}$\footnote{Corresponding author, E-mail:~wuguocheng2002@yahoo.com.cn (G.C. Wu)}
\vspace{4mm}, Sheng Zhang$^{2}$\\
1. Colledge of Textile, Donghua University, Shanghai 201620, P.R. China;\\
2. School of Mathematical Sciences, Dalian University of Technology, \\Dalian 116024, P.R. China.}
\date{}
\maketitle
--------------------------------------------------------------------------------------------------------------------\\
\leftline{\bf\ Abstract}

With the modified Riemann-Liouville fractional derivative, a fractional Tu
formula is presented to investigate generalized Hamilton structure
of fractional soliton equations. The obtained results can be reduced to the
classical Hamilton hierachy of ordinary calculus.

\vspace{0.5cm} \noindent{\bf Key words:}\quad  Fractionalized Tu
formula; Fractional Hamilton system; Fractional evolutionary
equations

\vspace{0.5cm} \noindent{\bf PACS:}\quad  02. 03. Ik; 05.45.Df; 05.30.Pr

--------------------------------------------------------------------------------------------------------------------\\
\section{Introduction}
Nobel Laureate Gerardus't Hooft once remarked that discrete
space-time is the most radical and logical viewpoint of reality. In
such discontinuous space-time, fractional calculus plays an important
role which can accurately describe many nonlinear phenomena in
physics, i.e., Brownian motion, anomalous diffusion, transportation
in porous media, chaotic dynamics, physical kinetics and quantum
mechanics.

Past decades witness the development of fractional calculus in
various fields, such as rheology, quantitative biology,
electrochemistry, scattering theory, diffusion, transport theory,
probability potential theory and elasticity. For details, see the
monographs of Kilbas et al. [1], Kiryakova [2], Lakshmikantham and
Vatsala [3], Miller and Ross [4] , and Podlubny [5].

Since Riewe [6] proposed a concept of non-conservation mechanics,
fractional conservation laws [7], fractional Lie symmetries [8] and
fractional Hamilton systems [9--16] have caught much attention.
 In recent study, Fujioka et al found that the propagation of optical
solitons can be described by an extended nonlinear
Shr$\ddot{o}$dinger equation which incorporates fractional
derivatives [17, 18].

 Searching for new integrable
hierarchies of soliton equations is an important and interesting
topic in soliton theory.
Tu scheme of ordinary calculus [19] is an efficient method to
generate integrable Hamilton systems. It took various efficient approaches to
have obtained many integrable systems such as AKNS hierarchy, KN
hierarchy, Schr$\ddot{o}$dinger system, and so on [20--26]. In order
to  consider the Hamilton structure of fractional soliton equations,
some questions may naturally arise: (1) Can we have a generalized Tu
sheme in fractional case? (2) How to define Hamilton equations for fractional soliton
hierachy?

In this study, we start from a Lax pair of fractional order in the
sense of the modified Riemann-Liouville's derivative [14] and propose a generalized
Tu sheme to investigate the Hamilton structure of fractional soliton
evolutionary equations.

\section{Modified Riemann-Liouville derivative}
Generally, there are two kinds of fractional derivatives: local
fractional derivatives and nonlocal ones.
 The most used nonlocal operator is the Caputo
derivative which requires the defined functions should be
differentiable. The condition is so strict that many engineering
problems cannot satisfy, i.e., functions defined on fractal curves,
fractional diffusion problem. As a result, the Caputo derivative is
not suitable for such problems theoretically.

Several local versions have been proposed: Kolwankar-Gangal's local
fractional derivative [27--29], Chen's fractal derivative [30, 31],
Cresson's derivative [32], Jumrie's modified Riemman-Liouville
derivative [33] and Parvate's $F^{\alpha}$ derivative [34], among which
Jumari's modified R-L derivative is defined as

\begin{equation}
D_x^\alpha  f(x) = \frac{1}{{\Gamma (1-\alpha )}}\int_0^x  (x - \xi
)^{-\alpha } (f(\xi )-f(0))d\xi ,~~0 < \alpha <1.
\label{eq5}
\end{equation}
Here the derivative on the right-hand side is the Riemann-Liouville
fractional derivative. The nonlinear techniques for such fractional
differentiable equations can be found in Refs. [8, 35, 36]

We can have
following results for Jumarie's modified Riemann-Liouville (R-L) derivative.

(a)  The Leibniz product law

If $f(x)$ is an $\alpha$ order differentiable function in the area of point $x$, from the Rolle-Kolwankar-Jumarie's Taylor series [31],
one can have

\begin{equation}
D_x^\alpha f(x)= \mathop {\lim }\limits_{y \to x }
\frac{\Gamma(1+\alpha)(f(y) - f(x ))}{{(y - x)^\alpha }},\;\;0 <
\alpha < 1. \label{eq2}
%(2)
\end{equation}

Assuming $g(x)$ is an $\alpha$ order differentiable function,
the Leibniz product law can hold

\begin{equation}D^{\alpha}_{x}(f(x)g(x))=g(x) D^{\alpha}_{x}f(x)+f(x)D^{\alpha}_{x}g(x). \end{equation}

(b) Integration with respect to $(dx)^\alpha $(Lemma\textbf{ 2.1} of
[37])

\begin{equation}
_0 I_x^\alpha  f(x) = \frac{1}{{\Gamma (\alpha )}}\int_0^x  (x - \xi
)^{\alpha  - 1} f(\xi )d\xi  = \frac{1}{{\Gamma (\alpha  +
1)}}\int_0^x
 f(\xi )(d\xi )^\alpha  ,0 < \alpha  \le 1.
\label{eq5}
%(5)
\end{equation}

\vskip 10pt

 (c) Generalized Newton-Leibniz Formulation

Assume $D^{\alpha}_{x}f(x)$ is a integrable function in the interval
$[a, b]$. Obviously,

\begin{equation}
\frac{1}{\Gamma(1+\alpha)}\int^{b}_{a}D^{\alpha}_{x}f(x)(dx)^{\alpha}=f(b)-f(a),
\end{equation}

\begin{equation}
\frac{1}{\Gamma(1+\alpha)}\int^{x}_{a}D^{\alpha}_{\xi}f(\xi)(d\xi)^{\alpha}=f(x)-f(a),
\end{equation}
and
\begin{equation}
\frac{D^{\alpha}_{x}}{\Gamma(1+\alpha)}\int^{x}_{a}f(\xi)(d\xi)^{\alpha}=f(x).
\end{equation}

(d) Integration by parts

With the properties (b) and (c), integration by parts for $\alpha$
order differentiable functions $f(x)$ and $g(x)$ can be presented as
\begin{equation}\frac{1}{\Gamma(1+\alpha)}\int^{b}_{a}g(x)D^{\alpha}_{x}f(x)(dx)^{\alpha}
=g(x)f(x)\mid^{b}_{a}-\frac{1}{\Gamma(1+\alpha)}\int^{b}_{a}f(x)D^{\alpha}_{x}g(x)(dx)^{\alpha}.
\end{equation}

The above properties (a)--(d) can be found in Ref. [33].

 \vskip 10pt (e) Fractional variational derivative

From Jumarie's variational derivative [14] and Almeida' fractional variational approach [15], the
fractional variational derivative is defined as
\begin{equation}\frac{\delta L}{\delta y}=\frac{{\partial L}}{{\partial y}} +\sum_{k=1}(-1)^{k}  ~(D_x^\alpha)^{k} (\frac{{\partial
L}}{{\partial (D_x^\alpha)^{k}  y}}),\end{equation}
where $k$ is a positive
integer.

(f) From Eq. (2), we can have
\begin{equation}
D_x^\alpha f(x)= \mathop {\lim }\limits_{h \to 0 }
\frac{\Gamma(1+\alpha)(f(x+h) - f(x))}{{h^\alpha
}}=\frac{\Gamma(1+\alpha)df(x)}{(dx)^{\alpha}},\;\;0 < \alpha < 1.
\label{eq2}
%(2)
\end{equation}
As a result, we can find that
\begin{equation}
f(b)-f(a)=\frac{1}{\Gamma(1+\alpha)}\int^{b}_{a}D^{\alpha}_{x}f(x)(dx)^{\alpha}=\frac{1}{\Gamma(1+\alpha)}\int^{b}_{a}d^{\alpha}f(x).
\end{equation}

\vskip 10pt

\section{Fractional Hamilton Structure}

A number of useful attempts have been made to establish fractional
variational principles and Hamilton system [9--16]. Different types
of fractional derivatives may lead to different results, for
examples, i.e., Baleanu's fractional Hamilton system with Caputo
derivative [10], Riemann-Liouville type Hamilton mechanics [11],
Argwal's Hamilton Formulation with Riesz derivative [12] and
Jumairie's Lagrange formula [14]. In this section, we revisit
Jumarie's fractional Hamilton system.

\subsection{A Fractional Exterior Differential
Approach}

Since Ben Adda proposed the fractional generalization of
differential [38, 39], many fractional exterior differential
approaches and applications related to different forms of fractional
derivatives appeared in open literature [40--42]. A brief review is
available in Tarasov's work [43].

Starting from the total derivative in the integer dimensional space,
and assuming \begin{math}f = f(u,v),\end{math}
\begin{math}u = u(x)\end{math} and \begin{math}v = v(x)\end{math},
where $u$, $v$ are $\alpha$ order differentiable fuctions and $f$ is a differentiable function with respect to $u$ and $v$,
we obtain the total derivative as follows
\begin{equation}df=f _u du +
 f _v dv.\end{equation}

Mutiple the both sides of Eq. (12) with $\Gamma(1+\alpha)$, we can get

\begin{equation}d^{\alpha}f=f _u d^{\alpha}u  +
 f _v d^{\alpha}v=f _u D^{\alpha}_{x}u(dx)^{\alpha}  +
 f _v D^{\alpha}_{x}v(dx)^{\alpha}.\end{equation}

On the other hand, if we assume that
 $u$, $v$ are differentiable fuctions and $f$ is a $\alpha$ order differentiable function with respect to $u$ and $v$

\begin{equation}df= \frac{f^{(\alpha)}_u}{\Gamma(1+\alpha)} (du)^{\alpha} +
 \frac{f^{(\alpha)}_v }{\Gamma(1+\alpha)}(dv)^{\alpha}{\rm{,}}\end{equation}
we can derive a different definition of $d^{\alpha}$ as follows

\begin{equation}d^{\alpha}f= f^{(\alpha)}_u (u_{x})^{\alpha}(dx)^{\alpha}
 + f^{(\alpha)}_v(v_{x})^{\alpha}(dx)^{\alpha}{\rm{.}}\end{equation}

Thus, we may have two different results for $D^{\alpha}_{x}x^2$,
$\frac{2 \Gamma(1+\alpha)x^{1-\alpha}}{\Gamma(2-\alpha)}$  or
$\frac{2 x^{2-\alpha}}{\Gamma(3-\alpha)}$, respectively, if we know
nothing about the differentiablity of $x^2$.

\subsection{Fractional Hamilton Equations}
We define the fractional functional

\begin{equation}J[p,q] = \frac{1}{\Gamma(1+\alpha)}\int {[pD_t^\alpha q -
H(t,p,q)]{\rm{(}}d} t)^\alpha
\end{equation}

Then, we can readily derive the fractional
Poincare\textendash{}Cartan 1-form, which reads

\begin{equation}\omega = pd^{\alpha}  q - H(dt)^\alpha. \end{equation}

From Eq. (17), we have

\begin{equation}
d^\alpha  \omega  = p_t ^{(\alpha )} (dt)^\alpha   \wedge d^\alpha q
+ d^\alpha  p \wedge d^\alpha  q - \frac{{\partial H}}{{\partial
p}}d^\alpha  p \wedge (dt)^\alpha   - \frac{{\partial H}}{{\partial
q}}d^\alpha  q \wedge (dt)^\alpha   $$$$ = {\rm{[}}p_t ^{(\alpha )}
+ \frac{{\partial H}}{{\partial q}}{\rm{]}}(dt)^\alpha \wedge
d^\alpha q + {\rm{[}}\frac{{\partial H}}{{\partial p}}(dt)^\alpha -
d^\alpha q{\rm{]}} \wedge d^\alpha  p{\rm{.}}
\end{equation}

The fractional closed condition \begin{math}d^\alpha  \omega  =
0\end{math} allows us to obtain the following fractional Hamilton
equations
\begin{equation}
D_t ^{{\rm{(}}\alpha {\rm{)}}} q = \frac{{\partial H}}{{\partial
p}}, \end{equation} and
\begin{equation} D_t ^{{\rm{(}}\alpha
{\rm{)}}} p =  - \frac{{\partial H}}{{\partial q}}{\rm{.}}
\end{equation}

We must point out, the results Eq. (13), Eq. (15), Eq. (19) and Eq. (20) can be found in Ref. [14].

\section{Fractional Tu Formula and Its Application}

\subsection{A Fractionalized Tu Formula}
Set$~A_{n}={A=(a_{i,~j}),~a_{i,~j}\in C}$. Assume $A$ and $B$ $\in
C$. Deffine $[A,~B]=AB-BA$. Hence, $A_{n}$ is a Lie algebra. The
corresponding loop algebra is defined as
\begin{equation}\tilde{A}_{n}={A(n)=A\lambda^{n},~n\in Z.}\end{equation}

 Tu Formula is a beautiful identity to generate
integral Hamilton equations. In the past decades, many integral
Hamilton hierarchies are obtained via this technical [20--26].
Consider the fractional compatibility condition,
\begin{equation}
\phi^{(\alpha)}_{x}(x, t)=U\phi,~~\phi^{(\beta)}_{t}(x, t)=V\phi,
\end{equation}
where the fractional derivative is in the sense of the modified R-L derivative [14, 31], and $\phi$ is a $n$-dimensional function vector.

The compatibility condition of Eq. (22) leades to the generalized zero
curvature equation
\begin{equation}
U^{(\beta)}_{t}-V^{(\alpha)}_{x}+[U,~V]=0, ~~[U,~V]=UV-VU.
\end{equation}

When taking $\alpha=\beta=1$, Eq. (22) reduces to the classical zero
curvature equation. Set
\begin{equation}
U=e_{0}(\lambda)+\sum^{n}_{i=1}e_{i}(\lambda)u_{i}, ~~
\{e_{i}(\lambda), 0\geq i \leq n \}\subset \tilde{A}_{n},
\end{equation}
where $u=u(u_{1}, u_{2},...,u_{n})^{T}$ denotes a vector function. By the
gradation of $\tilde{B}_{n}$, define rank$\lambda$=deg($\lambda$),
then  rank $(e_{0}(\lambda))=\alpha$,  rank
$(e_{i}(\lambda))=\alpha_{i}$, $0\leq i\leq n$ are all known. If we
take the ranks of $u_{i}$ as $\alpha-\alpha_{i}$, $1\leq i\leq n$,
then each term in $U$ is of the same rank $\alpha$, denoted by rank

\begin{equation} \rm{rank} ~(U)=rank~(\frac{\partial^{\alpha}}{\partial \rm{\emph{x}}^{\alpha}})=\alpha.\end{equation}

 If a solution of the stationary zero curvature equation

\begin{equation}
-V^{(\alpha)}_{x}+[U,~V]=0,
\end{equation}
is given by $V=\sum_{m\geq 0}V_{m}\lambda^{-m},
(V_{m})_{\lambda}=0$, $m\geq0$. rank $(V_{m})_{\lambda}$ is assumed
to be given so that rank $(V_{m})_{\lambda}=\beta, ~~m\geq0,$ then
each team in $V$ has the same rank, denoted by

\begin{equation}   \rm{rank}   (V)=rank(\frac{\partial^{\beta}}{\partial \emph{t}^{\beta}})=\eta.\end{equation}
 Suppose $f(A, B)=\rm{tr} (AB)$. The follwoing properties can
be satisfied

(a)~Symmetry relationship

 $$f(A, ~B)=f(B, ~A);$$

(b) The bilinearity can hold

$$f(c_{1}A_{1}+c_{2}A_{2},~B)=c_{1}f(A_{1}, B)+c_{2}f(A_{2},
~B);$$

(c) In the sense of the local fractional derivative, the gradient
$\nabla_{B}f(A,~B)$ of the functional $f(A,~B)$ is defined by
\begin{equation}
\frac{\partial}{\partial_{\epsilon}}f(A, ~B+\epsilon
C)=f(\delta_{B}f(A,~B),~C), \forall A, ~B,~C\in\tilde{ A}_{n},
\end{equation}
where $\delta_{B}$ is variational derivative with respect to $B$.

With the fractional variational derivative, we can have the
following results,
\begin{equation}
\delta_{B}f(A,~B^{k\alpha}_{x})=(-1)^{k}A^{(k\alpha)}_{x},
\end{equation}
where $k$ is a positive integer and
$D^{k\alpha}_{x}=\underbrace{D^{\alpha}_{x}...D^{\alpha}_{x}}_{k}$.

(d) Communication relationship
\begin{equation}
f([A,~B],~C)=f(A, [B,~C]),\forall A, ~B,~C\in\tilde{ A}_{n}.
\end{equation}

Construct a functional
\begin{equation}
W=f(V,~U_{\lambda})+f(K, V^{(\alpha)}_{x}-[U,~V]),
\end{equation}
where $U$, $V$ meet Eq. (22), $K\in \tilde{A}_{n}$, rank
$K$=-rank$~\lambda$.

With the defined fractional variational derivative,
\begin{equation}
\frac{\delta W}{\delta K}=V^{(\alpha)}_{x}-[U,~V],~~\frac{\delta
W}{\delta V}=U_{\lambda}-K^{(\alpha)}_{x}+[U,~V],
\end{equation}
from the above equations, we can derive

$$
[K,~V]^{(\alpha)}_{x}=[K^{(\alpha)}_{x},~V]+[K,~V^{(\alpha)}_{x}]~~~~~~~~~~~~~~~~~~~~~~~~~~~~~~$$$$
=[U_{\lambda}+[U,~K],~V]+[K,~[U,~V]]~~~~~
$$
$$=[U_{\lambda},~V]+[[U,~K],V]+[[V, U],~K]$$
$$=[U_{\lambda},~V]+[U,~[K,~V]].~~~~~~~~~~~~~~~~~$$

We can check $V^{'}=[K,~V]-V_{\lambda}$ satisfies Eq. (26) and
$\frac{V}{\lambda}$ also satisfies Eq. (26) since
rank$(Z)$=rank$(V_{\lambda})$=rank$(V)$-rank$(\lambda)$=rank$(\frac{V}{\lambda})$.
Therefore, if two solutions of Eq. (23), $V$ and $V^{'}$ are
linearly dependent, we can get $V^{'}=\frac{\gamma}{\lambda}V$.

Using Eq. (31) again, we can have a fractional trace identy as
follows

$$
 \frac{\delta f(V,~U_{\lambda})}{\delta u_{i}}=f(V,~\frac{\partial
U_{\lambda}}{\partial u_{i}})+f([K,~V], ~\frac{\partial
U_{\lambda}}{\partial
u_{i}})~~~~~~~~~~~~~~~~~~~~~~~~~~~~~~~~~~~~~~~~~~~
$$
$$=f(V,~\frac{\partial U_{\lambda}}{\partial
u_{i}})+f(V_{\lambda}, ~\frac{\partial U_{\lambda}}{\partial
u_{i}})+\frac{\gamma}{\lambda}f(V,~\frac{\partial U}{\partial
u_{i}})~~~~~~~~~~~~~~~~~$$
$$=\frac{\partial}{\partial\lambda}f(V,~\frac{\partial U}{\partial
u_{i}})+f(V_{\lambda}, ~\frac{\partial U_{\lambda}}{\partial
u_{i}})+(\lambda^{-\gamma}\frac{\partial}{\partial\lambda}\lambda^{\gamma})f(V,~\frac{\partial
U}{\partial u_{i}})~$$
$$=\lambda^{-\gamma}\frac{\partial}{\partial\lambda}[\lambda^{\gamma}f(V,~\frac{\partial U}{\partial
u_{i}})],~~0\leq i\leq n.~~~~~~~~~~~~~~~~~~~~~~~~~~$$

We must point out, the variational derivative here is defined in the sense of the modified R-L
fractional derivative.

\subsection{Fractional Soliton Hierarchies and Their Hamilton Structures}

Recently, Fujioka et al [17] found the propagation of optical
solitons can be described by an extended NLS equation which
incorporates fractional derivatives. The detailed review can be
found in Ref. [18]. In view of this point, we consider fractional
AKNS hierarchy strating from the generalized spectral problem
\begin{equation}
\Phi^{(\alpha)}_x=U(\lambda,u)=\left(
\begin{array}{cc}
-\lambda&q\\
 r& \lambda
\end{array}
\right)\Phi,~~u=\left(
\begin{array}{c}
q\\
\\r
\end{array}
\right),~~\Phi=\left(
\begin{array}{c}
\phi_1\\
\phi_2
\end{array}
\right).
\end{equation}
Choose a simple subalgebra of $A_{1}$

\begin{equation}
e_{1}(0)=\left(
\begin{array}{cc}
1&0\\
 0& -1
\end{array}
\right),~e_{2}(0)=\left(
\begin{array}{cc}
0&1\\
 0& 0
\end{array}
\right),~e_{3}(0)=\left(
\begin{array}{cc}
0&0\\
 1& 0
\end{array}
\right),
\end{equation}
equipped with the commutative relations
\begin{equation}
[e_{1}(m),~e_{2}(n)]=2e_{2}(m+n),~[e_{1}(m),~e_{3}(n)]=-2e_{3}(m+n),~[e_{2}(m),~e_{3}(n)]=e_{1}(m+n).
\end{equation}
Then, we find that the adjoint representation equation
$V^{(\alpha)}_x=[U,V]=UV-VU$ yields

$$ a^{(\alpha)}_{0_x}=qc_0-rb_0, b_0=0, c_0=0,$$$$
a^{(\alpha)}_{ix}=qc_i-rb_i,$$
$$b^{(\alpha)}_{ix}=-2b_{i+1}-2qa_{i},$$
$$c^{(\alpha)}_{ix}=2ra_{i}+2c_{i+1},i\geq1,$$

the first few of which reads $$a_0=-1, ~b_0=0, ~c_0=0,$$
$$a_1=0,~b_1=q, ~c_1=r,~b_2=-\frac{1}{2}q^{(\alpha)}_x,$$
$$c_2=\frac{1}{2}r^{(\alpha)}_x,~a_2=\frac{1}{2}qr.$$

We  can derive the recurenece relationship

 \begin{equation} \left(
\begin{array}{c}
c_{n+1}\\b_{n+1}
\end{array}\right)
=\frac{1}{2}\left(
\begin{array}{cc}
D^{\alpha}_{x}-2rD^{-\alpha}_{x}q&2rD^{-\alpha}_{x}r\\
-2qD^{-\alpha}_{x}q&-D^{\alpha}_{x}+2qD^{-\alpha}_{x}
\end{array}
\right) \left(
\begin{array}{c}
c_n\\b_n
\end{array}
\right)= L\left(
\begin{array}{c}
c_n\\b_n
\end{array}
\right),
\end{equation}
\[
a_n=D^{-\alpha}_{x}(qc_n-rb_n), n=0,1,2,3\cdots.
\]

Denoting

$$(V_+^{(n)})^{(\alpha)}_{x
}=\sum_{i=0}^n a(i)e_{1}(n-i)+b(i)
e_{2}(n-i)+c(i)e_{3}(n-i),$$
$$V_-^{(n)}=\lambda^{n}V-V_+^{(n)},$$
Eq. (26) can be written as
\begin{equation}-(V_+^{(n)})^{(\alpha)}_{x
}+[U,V_+^{(n)}]=(V_-^{(n)})^{(\alpha)}_{x
}-[U,V_-^{(n)}].\end{equation} It is easy to verify that the terms
on the left-hand side in $(37)$ are of degree $\geq 0$, while the
terms on the right-hand side in Eq. (37) are of degree $\leq 0$. Thus,
we have
$$-(V_+^{(n)})^{(\alpha)}_{x
}+[U,V_+^{(n)}]=2b(n+1)e_{2}(0) -2c(n+1)e_{3}(0).$$

Taking an arbitrary modified term for $V_+^{(n)}$ as
$\bigtriangleup_n=0$. Notice $V^{(n)}=V_+^{(n)}$, it is easy to
compute, the zero curvature equation
 \begin{equation}U^{(\beta)}_t-(V^{(n)})^{(\alpha)}_x+[U,V^{(n)}]=0, \end{equation}which
gives rise to

 \begin{equation}u^{\beta}_t=\pmatrix{q^{(\beta)}\cr r^{(\beta)}}_t=\pmatrix{0&-2\cr 2&0}\pmatrix{c(n+1)\cr
b(n+1)}=JL \pmatrix{c(n)\cr b(n)}=JL^n \left(
\begin{array}{c}
 r\\q
\end{array}
\right).  \end{equation} Here $J$
 is Hamiltonian operator.

For $n=2$, we obtain the fractional AKNS equations
\begin{equation}\label{akns2}
\left\{
\begin{array}{l}
D^{\beta}_{t}q_{t_1}=-\frac{1}{2}D^{\alpha}_{x}D^{\alpha}_{x}q+q^2 r, \\
D^{\beta}_{t}r_{t_1}=\frac{1}{2}D^{\alpha}_{x}D^{\alpha}_{x}r- qr^2.
\end{array}\right.
\end{equation}
When $\alpha=\beta=1$,  Eq. (40) can be reduced to the classical
AKNS system,\begin{equation}\label{akns2} \left\{
\begin{array}{l}
q_{t_1}=-\frac{1}{2}q_{xx}+q^2 r, \\
r_{t_1}=\frac{1}{2}r_{xx}- qr^2.
\end{array}\right.
\end{equation}

In order to use the proposed trace identity, a deirect compute leads
to

\begin{equation}
f(V,~U_{\lambda})=-2a_{n}, f(V,~\frac{\partial U}{\partial
q})=c_{n},~ f(V,~\frac{\partial U}{\partial r})=b_{n}.
\end{equation}
As a result,
\begin{equation}
{\frac{\delta(-2a)}{\delta u}}=\lambda^{-\gamma }\frac{\partial
}{\partial\lambda}\lambda^{\gamma}\pmatrix{c\cr b}.
\end{equation}

Compared with the coefficients of $\lambda^{-n-1}$,

\begin{equation}
{\frac{\delta(-2a_{n+1})}{\delta u}}= (\gamma-n)\pmatrix{c_{n}\cr
b_{n}}.
\end{equation}
Setting $n=1$, we can determine $\gamma=0$ from the initial values.
Then we can derive the fractional Hamilton function

\begin{equation}
H_n=\frac{2a_{n+1}}{n},~~\frac{\delta H_n}{\delta
u}=\pmatrix{c_{n}\cr b_{n}}.
\end{equation}
The generalized evolutionary equations can be given as

\begin{equation}
\label{5} u^{(\beta)}_{t_n}=\left(\begin{array}{c}
q^{(\beta)}_{t_n}\\r^{(\beta)}_{t_n}
\end{array}\right)
 =J\frac{\delta H_n}{\delta u}.
\end{equation}

\section*{Acknowledgments}

 \ \ \ \
The first author feels grateful to Dr. Xiao-Jun Yang for his helpful
discussion about the physical meaning of the local fractional
derivative and its possible use in other fields.

\section{Conclusion}

Inspired by the previous work [44], in this study we use a
different fractional derivative, modified Riemann-Liouville derivative,
establish a fractionalized Tu sheme for fractional
differential equations and define a local fractional Hamilton system
and derive fractional evolutionary soliton hierachies.

However,
there are still other interesting questions needed to be addressed
i.e., physical mearning of fractional soliton which may be related
to fractal media, fractional integral coupling method, nonlinear
techniques for fractional soliton equations. Such work is under
consideration.

%\end{CJK}
\end{document}